\documentclass{llncs}

\usepackage{amsfonts,epsfig}
\usepackage{hyperref}
\usepackage{array}
\usepackage{float}

\usepackage[left=2.5cm,right=2.5cm,top=3cm,bottom=3cm]{geometry}

\begin{document}

\title{EMFET: E-mail Features Extraction Tool}
\titlerunning{Hamiltonian Mechanics}  
%
\author{Wadi' Hijawi\inst{1,2} \and Hossam Faris\inst{1}
Ja'far Alqatawna\inst{1,2} \and Ibrahim Aljarah\inst{1} \and \\ Ala' M. Al-Zoubi\inst{1,2} \and Maria Habib\inst{1} }
%
%
%
\institute{King Abdullah II School for Information Technology\\
The University of Jordan, Amman, Jordan\\
\and
Jordan Information Security and Digital Forensics \\ Research Group (JISDF) 
Amman, Jordan \\
\email{w.hijjawi@jisdf.org,\{hossam.faris, j.alqatawna, i.aljarah\}@ju.edu.jo, \\ alaah14@gmail.com,maryahabeeb@yahoo.com},
}


\maketitle              

\begin{abstract}

EMFET is an open source and flexible tool that can be used to extract a large number of features from any email corpus with emails saved in EML format. The extracted features can be categorized into three main groups: header features, payload (body) features, and attachment features. The purpose of the tool is to help practitioners and researchers to build datasets that can be used for training machine learning models for spam detection. So far, 140 features can be extracted using EMFET. EMFET is extensible and easy to use. The source code of EMFET is publicly available at GitHub

(\url{https://github.com/WadeaHijjawi/EmailFeaturesExtraction})

\end{abstract}

\section{Interface}

\begin{figure*}[htb]
\centering
\includegraphics[scale=0.62]{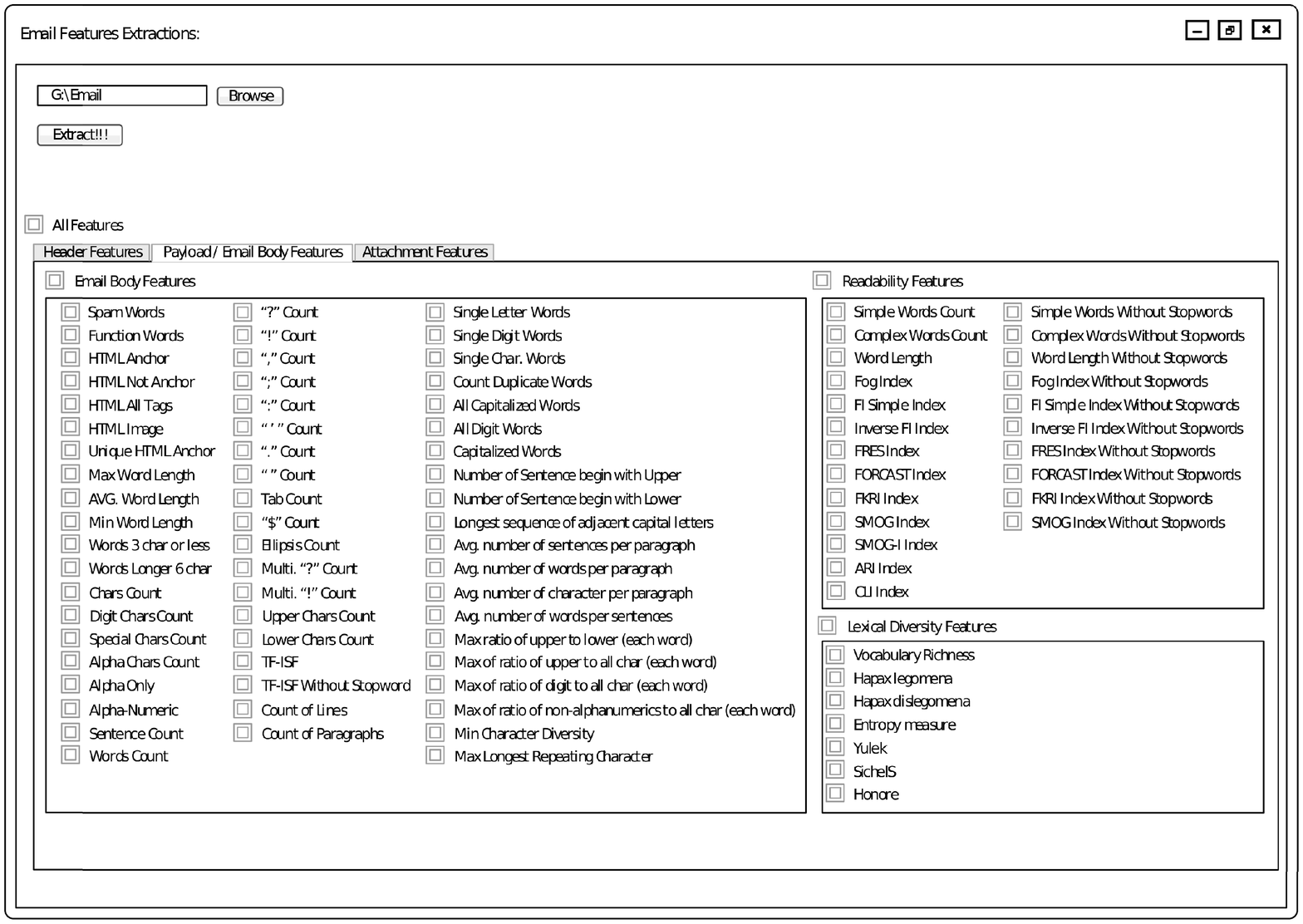}
\caption{Features Extraction Tool interface.}
\label{fig:interface}
\end{figure*}

The interface of EMFET is designed as a group of tab pages. Each tab page represents a feature category and contains checkboxes with unique tags for selecting the desired features to extract. Additionally, the interface provides the user a folder browser to select the corpus folder from which the features will be extracted as shown in Fig \ref{fig:interface}. A user-friendly progress bar is used to view the current status of the extraction process.

\section{Functionality}

In EMFET, the user can load an email corpus with emails saved in EML format (e.g. SpamAssassin\footnote{SpamAssassin corpus. Available at:\url{http://spamassassin.apache.org/downloads.cgi?update=201504291720}}, CSDMC2010\footnote{CSDMC2010 corpus. Available at:\url{http://csmining.org/index.php/spam-email-datasets-.html}} and Ling-Spam\footnote{Ling-Spam corpus. Available at:\url{http://www.csmining.org/index.php/ling-spam-datasets.html}}). EML file extension for an email message saved to a file in the Internet Message Format protocol for electronic mail messages. EML is one of the most common file extensions used by email applications such as Outlook, Thunderbird and Gmail. Several public email corpora in EML format are available on the web and can be used with EMFET.

The tool splits the emails' into their main parts which are: \texttt{From}, \texttt{To}, \texttt{CC}, \texttt{BCC}, \texttt{Subject}, \texttt{Text Body}, and \texttt{HTML Body} and save each of them to an output file. Then the tool extracts the selected features from the corresponding email part and save them in another file in CSV format. If any error occurred during the extraction process, an error file will be generated.

\begin{figure}[h]
\begin{center}
\includegraphics[scale=0.5]{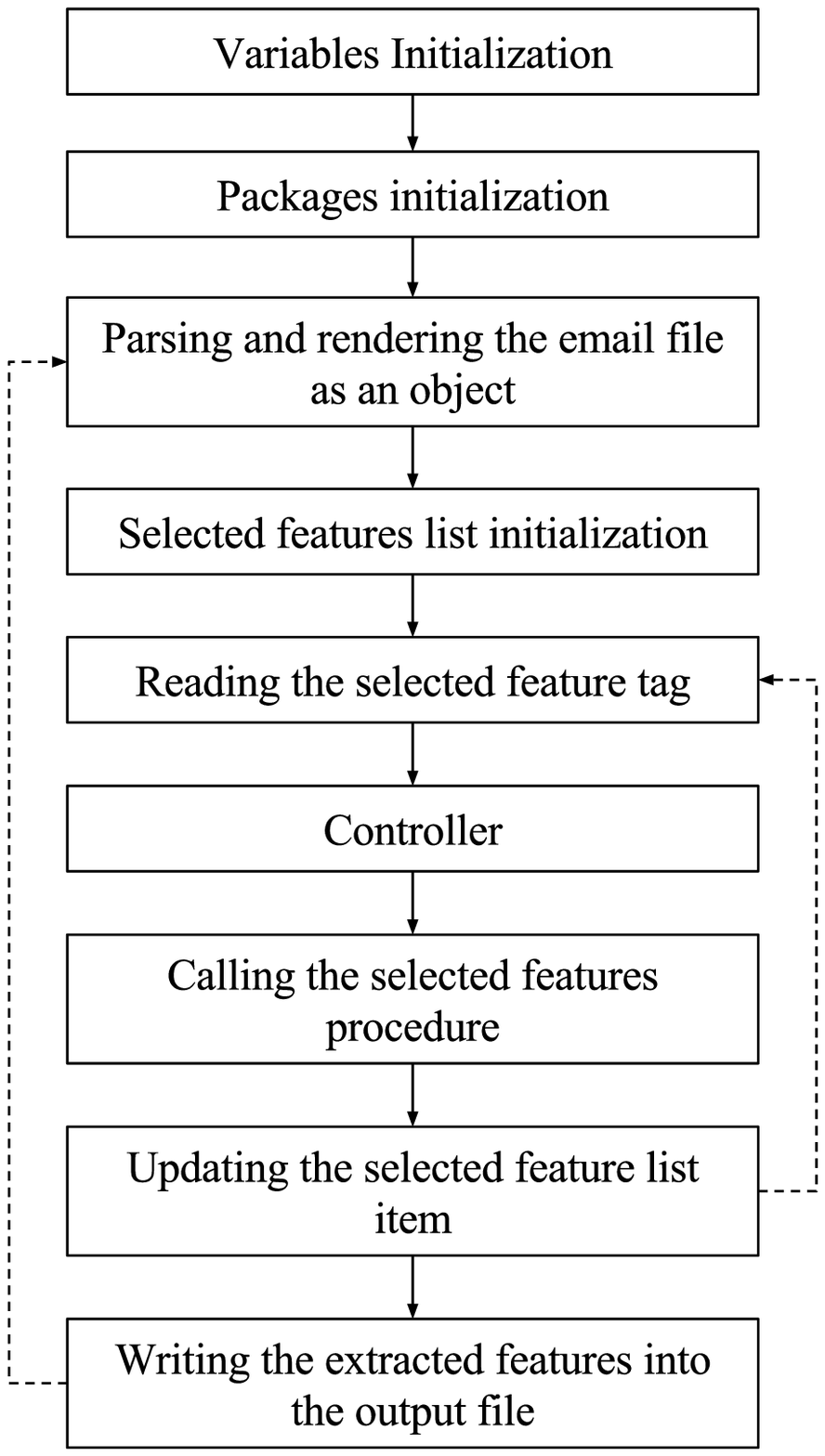}
\caption{The implementation structure of EMFET.}
\label{fig:imp}
\end{center}
\end{figure}

\section{Implementation}

The source code for features extraction process is organized in one main class with a controller to call each selected feature's procedure, which will use each email object's attributes and methods to extract the desired features. The variables, packages and emails objects are initialized in separated classes. Figure \ref{fig:imp} describes the implementation structure.

Different external libraries are utilized in the implementation which include: Microsoft CDO for Windows 2000 COM Library to deal with each email file as an object and save the emails' parts into an output file. For features extraction development, we used the following libraries:

\begin{itemize}
   \item HTML Agility Pack\footnote{External package to parse HTML. Available at:\url{https://htmlagilitypack.codeplex.com/}}.
   
   \item IKVM.NET\footnote{External package to enable Java and .NET interoperability. Available at:\url{https://www.ikvm.net/}}.
       
    \item Stanford.NLP.CoreNLP\footnote{External package to provides a set of natural language analysis tools. Available at:\url{https://www.nuget.org/packages/Stanford.NLP.CoreNLP/}}.   
   
   \item RegularExpressions\footnote{Internal package to work with text such as search a pattern within text.}.

   \item LINQ (.NET Language-Integrated Query)\footnote{Internal package to query and access data in objects such as XML documents, databases.}.
\end{itemize}

The simplified design and implementation structure of the tool makes it easy to use and enhances the ability to add more features to the tool later. More features can be added by creating a checkbox with a unique tag for the new feature, adding the extraction procedure in the main class and linking between this procedure's name and checkbox's tag in the controller. Additionally, passing parameters for each email as an object with a unique tag for each feature into the extraction procedures, this allows dealing with this procedures as independent problems. Moreover, using the internal and external packages reduces the source code lines and make the implementation easier to extend and modify.

\section{List of E-mail features}

\begin{figure*}[htb]
\centering
\includegraphics[scale=0.85]{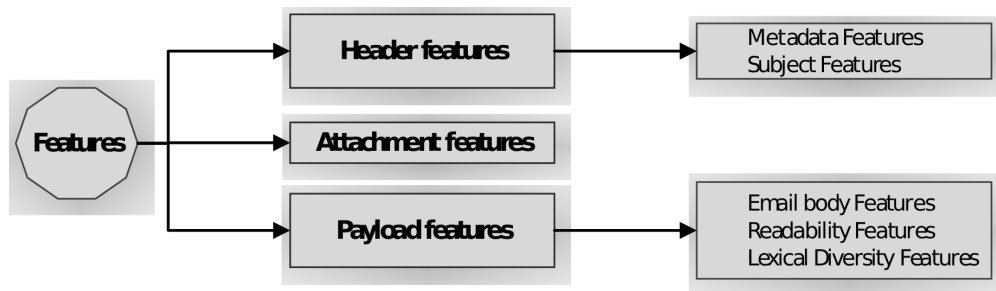}
\caption{Hierarchy of features groups.}
\label{fig:Hierarchy}
\end{figure*}

Based on the related studies in the literature, the spam features can be classified into three main categories: header features, payload (body) features, and attachment features \cite{wa2017imp}. Figure \ref{fig:Hierarchy} shows the hierarchy of the features categories. So far there are 140 features implemented in EMFET (i.e 49 header features, 2 attachment features, and 89 Payload features). Each feature category is described as follows.



\subsection{Header features} 
The email header is an essential element of any email message, which consists of a set of important features that helps in email delivery. These features are grouped into two classes, namely; email's metadata and subject. Table \ref{headerF} lists the header features included in TEMFET with brief descriptions.

\begin{table*}[http]
\centering
\caption{THE HEADER FEATURES}
 \scalebox{0.82}{
\begin{tabular}{|l|l|l|l|l|>
{\centering}l|l|l|}
\hline 

ID & Feature Details & Type & Ref. & ID & Feature Details & Type & Ref.\tabularnewline
\hline 
\hline 
1 & Year & Metadata & \cite{alqatawna2015improving} & 26 & Replay to MIL? & Metadata & \cite{alqatawna2015improving}\tabularnewline
\hline 
2 & Month & Metadata & \cite{alqatawna2015improving} & 27 & Replay to Yahoo? & Metadata & \cite{alqatawna2015improving}\tabularnewline
\hline 
3 & Day & Metadata & \cite{tran2013towards,alqatawna2015improving} & 28 & Replay to AOL? & Metadata & \cite{alqatawna2015improving}\tabularnewline
\hline 
4 & Hour & Metadata & \cite{tran2013towards,alqatawna2015improving} & 29 & Replay to Gov? & Metadata & \cite{alqatawna2015improving}\tabularnewline
\hline 
5 & Minute & Metadata & \cite{tran2013towards,alqatawna2015improving} & 30 & X-Mailman-Version & Metadata & \cite{alqatawna2015improving}\tabularnewline
\hline 
6 & Second & Metadata & \cite{tran2013towards,alqatawna2015improving} & 31 & Exist Text/Plain? & Metadata & \cite{alqatawna2015improving}\tabularnewline
\hline 
7 & From Google? & Metadata & \cite{alqatawna2015improving} & 32 & Exist Multipart/Mixed? & Metadata & \cite{alqatawna2015improving}\tabularnewline
\hline 
8 & From AOL? & Metadata & \cite{alqatawna2015improving} & 33 & Exist Multipart/Alternative? & Metadata & \cite{alqatawna2015improving}\tabularnewline
\hline 
9 & From Gov? & Metadata & \cite{alqatawna2015improving} & 34 & No. of characters. & Subject & \cite{tran2013towards}\tabularnewline
\hline 
10 & From HTML? & Metadata & \cite{alqatawna2015improving} & 35 & No. of capitalised words. & Subject & \cite{tran2013towards}\tabularnewline
\hline 
11 & From MIL? & Metadata & \cite{alqatawna2015improving} & 36 & No. of words in all uppercase. & Subject & \cite{tran2013towards}\tabularnewline
\hline 
12 & From Yahoo? & Metadata & \cite{alqatawna2015improving} & 37 & No. of words that are digits. & Subject & \cite{tran2013towards}\tabularnewline
\hline 
13 & From Example? & Metadata & \cite{alqatawna2015improving} & 38 & No. of words containing only letters. & Subject & \cite{tran2013towards}\tabularnewline
\hline 
14 & To Hotmail? & Metadata & \cite{alqatawna2015improving} & 39 & No. of words containing letters and numbers. & Subject & \cite{tran2013towards}\tabularnewline
\hline 
15 & To Yahoo? & Metadata & \cite{alqatawna2015improving} & 40 & No. of words that are single letters. & Subject & \cite{tran2013towards}\tabularnewline
\hline 
16 & To Example? & Metadata & \cite{alqatawna2015improving} & 41 & No. of words that are single digits. & Subject & \cite{tran2013towards}\tabularnewline
\hline 
17 & To MSN? & Metadata & \cite{alqatawna2015improving} & 42 & No. of words that are single characters. & Subject & \cite{tran2013towards}\tabularnewline
\hline 
18 & To Localhost? & Metadata & \cite{alqatawna2015improving} & 43 & Max ratio of uppercase to lowercase letters of each word  & Subject & \cite{tran2013towards}\tabularnewline
\hline 
19 & To Google? & Metadata & \cite{alqatawna2015improving} & 44 & Min character diversity of each word. & Subject & \cite{tran2013towards}\tabularnewline
\hline 
20 & To AOL? & Metadata & \cite{alqatawna2015improving} & 45 & Max ratio of uppercase letters to all characters of each word.  & Subject & \cite{tran2013towards}\tabularnewline
\hline 
21 & To Gov? & Metadata & \cite{alqatawna2015improving} & 46 & Max ratio of digit characters to all characters of each word.  & Subject & \cite{tran2013towards}\tabularnewline
\hline 
22 & To MIL? & Metadata & \cite{alqatawna2015improving} & 47 & Max ratio of non-alphanumerics to all characters of each word  & Subject & \cite{tran2013towards}\tabularnewline
\hline 
23 & Count of \textquotedblleft To\textquotedblright{} Email & Metadata & \cite{tran2013towards} & 48 & Max of the longest repeated character. & Subject & \cite{tran2013towards}\tabularnewline
\hline 
24 & Replay to Google? & Metadata & \cite{alqatawna2015improving} & 49 & Max of the character lengths of words. & Subject & \cite{tran2013towards}\tabularnewline
\hline 
25 & Replay to Hotmail? & Metadata & \cite{alqatawna2015improving} & - & - & - & -\tabularnewline
\hline 

\end{tabular}}
\label{headerF}
\end{table*}

\subsection{Payload features}
These features are categorized into three sub categories: Email body features, Readability features, and Lexical diversity features.

\subsubsection{Email body features} The e-mail body contains unstructured data such as images, HTML tags, text, and other objects. 
This features set contains 59 features, which are described in  Table \ref{bodyF}.

\begin{table*}
\centering
\caption{EMAIL BODY FEATURES}
\scalebox{0.84}{
\begin{tabular}{|l|l|l|l|l|l|l|}
\hline 
ID & Feature Details & Ref. & ID & Feature Details& Studies\tabularnewline
\hline 
\hline 
1 & Count of Spam Words & \cite{shams2016supervised,alqatawna2015improving,al2016voting}  & 31 & Number of question marks & \cite{alqatawna2015improving,al2016voting} \tabularnewline
\hline 
2 & Count of Function Words & \cite{shams2016supervised,al2016voting} & 32 & No. of multiple question marks & \cite{alqatawna2015improving}\tabularnewline
\hline 
3 & Count of HTML Anchor & \cite{tran2013towards,shams2016supervised,alqatawna2015improving,al2016voting} & 33 & No. of exclamation marks & \cite{alqatawna2015improving,al2016voting}\tabularnewline
\hline 
4 & Count of Unique HTML Anchor & \cite{tran2013towards,alqatawna2015improving} & 34 & No. of multiple exclamation marks & \cite{alqatawna2015improving}\tabularnewline
\hline 
5 & Count of HTML Not Anchor & \cite{shams2016supervised} & 35 & No. of colons & \cite{alqatawna2015improving,al2016voting} \tabularnewline
\hline 
6 & Count of HTML Image & \cite{al2016voting} & 36 & No. of ellipsis & \cite{alqatawna2015improving,al2016voting}\tabularnewline
\hline 
7 & Count of HTML All Tags & \cite{shams2016supervised,al2016voting} & 37 & Total No. of sentences & \cite{shams2016supervised,alqatawna2015improving,al2016voting} \tabularnewline
\hline 
8 & Count of Alpha-numeric Words & \cite{tran2013towards,shams2016supervised,al2016voting} & 38 & Total No. of paragraphs & \cite{alqatawna2015improving}\tabularnewline
\hline 
9 & TF-ISF & \cite{shams2016supervised} & 39 & Average No. of sentences per paragraph & \cite{alqatawna2015improving}\tabularnewline
\hline 
10 & TF\textbullet ISF without stopwords & \cite{shams2016supervised} & 40 & Average number of words pre paragraph & \cite{alqatawna2015improving}\tabularnewline
\hline 
11 & Count of duplicate words. & \cite{al2016voting} & 41 & Average No. of character per paragraph & \cite{alqatawna2015improving}\tabularnewline
\hline 
12 & Minimum word length & \cite{al2016voting} & 42 & Average No. of word per sentences & \cite{alqatawna2015improving}\tabularnewline
\hline 
13 & Count of lowercase letters & \cite{al2016voting} & 43 & No. of sentence begin with upper case & \cite{alqatawna2015improving}\tabularnewline
\hline 
14 & Longest sequence of adjacent capital letters & \cite{alqatawna2015improving,al2016voting}  & 44 & No. of sentence begin with lower case & \cite{alqatawna2015improving}\tabularnewline
\hline 
15 & Count of lines & \cite{alqatawna2015improving,al2016voting}   & 45 & Character frequency \textquotedblleft \$\textquotedblright{} & \cite{alqatawna2015improving,al2016voting} \tabularnewline
\hline 
16 & Total No. of digit character & \cite{alqatawna2015improving} & 46 & No. of capitalized words. & \cite{tran2013towards}\tabularnewline
\hline 
17 & Total No. of white space & \cite{alqatawna2015improving} & 47 & No. of words in all uppercase. & \cite{tran2013towards}\tabularnewline
\hline 
18 & Total No. of upper case character & \cite{alqatawna2015improving,al2016voting} & 48 & Number of words that are digits. & \cite{tran2013towards}\tabularnewline
\hline 
19 & Total No. of characters & \cite{tran2013towards,alqatawna2015improving} & 49 & No. of words containing only letters. & \cite{tran2013towards}\tabularnewline
\hline 
20 & Total No. of tabs & \cite{alqatawna2015improving} & 50 & No. of words that are single letters. & \cite{tran2013towards}\tabularnewline
\hline 
21 & Total No. of special characters & \cite{alqatawna2015improving} & 51 & No. of words that are single digits. & \cite{tran2013towards}\tabularnewline
\hline 
22 & Total number of alpha characters & \cite{alqatawna2015improving} & 52 & Number of words that are single characters. & \cite{tran2013towards}\tabularnewline
\hline 
23 & Total No. of words & \cite{alqatawna2015improving,al2016voting} & 53 & Max ratio of uppercase letters to lowercase letters of each word. & \cite{tran2013towards}\tabularnewline
\hline 
24 & Average word length & \cite{alqatawna2015improving,al2016voting} & 54 & Min of character diversity of each word. & \cite{tran2013towards}\tabularnewline
\hline 
25 & Words longer than 6 characters & \cite{alqatawna2015improving} & 55 & Max ratio of uppercase letters to all characters of each word.  & \cite{tran2013towards}\tabularnewline
\hline 
26 & Total No. of words (1 - 3 Characters) & \cite{alqatawna2015improving} & 56 & Max ratio of digit characters to all characters of each word.  & \cite{tran2013towards}\tabularnewline
\hline 
27 & No. of single quotes & \cite{alqatawna2015improving,al2016voting} & 57 & Max ratio of non-alphanumerics to all characters of each
word.  & \cite{tran2013towards}\tabularnewline
\hline 
28 & No. of commas & \cite{alqatawna2015improving,al2016voting} & 58 & Max of the longest repeating character. & \cite{tran2013towards}\tabularnewline
\hline 
29 & No. of periods & \cite{alqatawna2015improving,al2016voting} & 59 & Max of the character lengths of words. & \cite{tran2013towards,al2016voting}\tabularnewline
\hline 
30 & No. of semi-colons & \cite{alqatawna2015improving,al2016voting} & - & - & - \tabularnewline
\hline 

\end{tabular}}
\label{bodyF}
\end{table*}


\subsubsection{Readability features}
Readability features represent the difficulty properties of reading a word, a sentence or a paragraph in the given email's body \cite{shams2016supervised},\cite{al2016voting}. Readability features are extracted based on the syllables --- sequence of speech sounds ---, which are used to distinguish between the simple and complex words. This set of features contains 23 features that measure the difficulty of reading the email's body. The readability features are discussed as follows:

\begin{itemize}
   \item Number of simple words features (with and without stopwords), such as each word has at most two syllables.
    
   \item Number of complex words features (with and without stopwords), such as each word contains three or more syllables

    \item Word length features (with and without stopwords) based on the number of syllables, and the number of the words in the text.
    
    \item Fog Index (FI) features (with and without stopwords), which are the most popular readability measure that used estimate the years of education experience to understand the text at glance.
   
    \item Flesch Reading Ease Score (FRES) features (with and without stopwords), which are used to asses the textual difficulty.
   
    \item SMOG index features (with and without stopwords), which are used to measure the difficulty of the text writing.

    \item FORCAST index features (with and without stopwords), which are used to measure the reading skills of the text that contains high percentage of simple words.

   \item Flesch-Kincaid Readability Index (FKRI) features (with and without stopwords), which are similar to the previous features, but with different weighting factors.

  \item  Simple Word FI features (with and without stopwords), which are similar to the Fog Index but with the simple words.
    
   \item  Inverse FI features (with and without stopwords).

  \item SMOG-I feature, which is used to measure the difficulty of the text writing.
   
  \item Automated Readability Index (ARI), which are used to measure the understandability of a given text.
  
  \item  Coleman-Liau Index (CLI), which is similar ARI, but with of syllables factor instead of characters factor.

\end{itemize}

\subsubsection{Lexical diversity features}

Lexical Diversity Features are extracted based on the vocabulary size in the text, where the word occurrences are counted using different constraints \cite{tran2013towards},\cite{tweedie1998variable}, \cite{choi2012finding}. This set of features contains seven features as follows: 

\begin{itemize}
  \item Vocabulary Richness, which represents the number of distinct words in a text.   
  \item Hapax legomena or V(1,N): represents the number of words occurring one time in the text with N words.
  \item Hapax dislegomena or V(2,N): represents the number of words occurring two times in the text with N words.
  \item Entropy measure, which asses the average amount of information in the text.

  \item YuleK, which is a text characteristic measure that is independent of text length.
  
  \item SichelS, which is the ratio of dislegomena to the number of distinct words.

  \item Honore, which is the ratio of hapax legomena (V (1,N)) to the vocabulary size N. 
\end{itemize}

\subsection{Attachment Features}
We use two features related to attachments: Number of all attachment files in an email and Number of unique content types of attachment files in an email.

\section{How to install and use}
\label{sec:installsteps}
In this section we list instructions for downloading, installing, and using EMFET version 1.0.0, as follows:
\begin{enumerate}
\item This tool runs over Windows operating system, either 32 bit or 64bit, and under .Net Framework 4.0 or above.
\item Make sure you have Internet connection and an Internet browser (such as Google chrome or Firefox).
\item Type in the search bar the address of the tool on Github as follows: 

(\url{https://github.com/WadeaHijjawi/EmailFeaturesExtraction})
\item In the middle of the page click "Clone or download" button, after that a popup menu will appear, then click on "Download ZIP" button.
\item Decompress the downloaded ZIP folder using archiving manager software such as WinRAR on windows.
\item Open the extracted folder and double click on "EmailFeaturesExtraction" folder, then open the bin folder and then the Debug folder. Lastly, double click the executable file "EmailFeaturesExtraction.exe".
\item A new window will appear, browse your corpus (make sure that it contains EML file extension). There are three tabs to choose the features among them, Header Features, Payload/ Email Body Features, and Attachment Features.
\item By clicking the "Extract" button, a new folder inside your corpus folder will be generated and contains a comma separated values (CSV) file of the extracted features.
\end{enumerate}

\section{Citing EMFET}

Hijawi, W., Faris, H., Alqatawna, J., Aljarah, I., Ala'M, A. Z., and M. Habib (2017). E-mail Features Extraction Tool. Available at: (\url{https://github.com/WadeaHijjawi/EmailFeaturesExtraction}).\\

\noindent W. Hijawi, H. Faris, J. Alqatawna, Ala'M, A. Z., and I. Aljarah, “Improving email spam detection
using content based feature engineering approach,” in Applied Electrical Engineering and Computing Technologies
(AEECT), IEEE, 2017.

\section{Acknowledgment}

This work has been supported in part by funded project by The University of Jordan.

\baselineskip 2.4ex
\bibliographystyle{ieeetr}
\bibliography{reference} 

\end{document}